# GONE WITH THE WIND ON_MARS (GOWON): A WIND-DRIVEN NETWORKED SYSTEM OF MOBILE SENSORS ON MARS

Faranak Davoodi[1, 3], Ali Hajimiri[2], Neil Murphy[1], Shouleh Nikzad[1], Issa Nesnas[1], Michael Mischna[1], Bill Nesmith[1], [1]Jet Propulsion Laboratory, [2]California Institute of Technology, Pasadena, California USA, [3]Contact PI: faranak.davoodi@jpl.nasa.gov

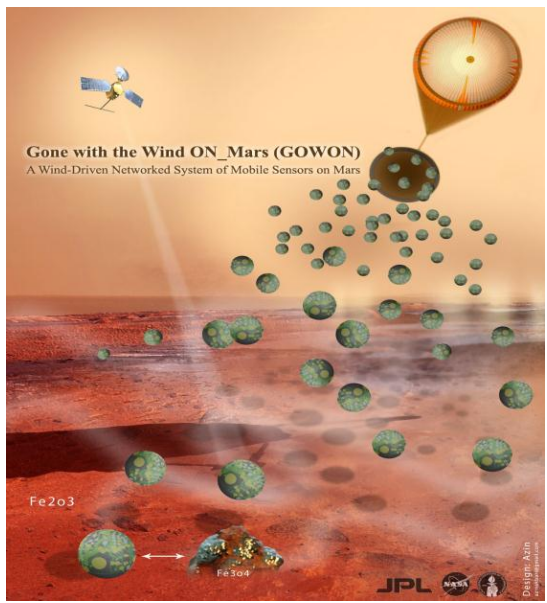

## Main Concept

We propose a revolutionary way of studying the surface of Mars using a wind-driven network of mobile sensors- **Go**ne with the **W**ind **ON**_Mars (**GOWON**). GOWON is envisioned to be a scalable, 100% self energy-generating and distributed system that allows in-situ mapping of a wide range of phenomena in a much larger portion of the surface of Mars compared to earlier missions. It could radically improve the possibility of finding rare phenomena like bio signatures through random wind-driven search. It could explore difficult terrains that were beyond the reach of previous missions, such as regions with very steep slopes, cluttered surfaces and/or sand dunes; GOWON is envisioned as an on going mission with a long life span. It could achieve any of NASA's scientific objectives on Mars in a cost-effective way, leaving a long lasting sensing and searching infrastructure on Mars.

GOWON would consist of a dynamic wireless network of many compact mobile sensors called *moballs*. The moballs are spherically shaped and wind-driven; they are lightweight and bouncy. (A calculation using typical wind speeds of 10-15m/s, obtained from the MarsWRF GCM[1], shows that for moballs ranging in mass from 140g (mass of an iPhone) to 500g, a radius of 35-75cm is sufficient to guarantee wind-driven motion.) The moballs perform in situ detection of essential environmental elements such as vaporized water, methane, wind, dust, clouds, light and UV exposure, temperature, as well as minerals of interest, possible bio-signatures, surface magnetic fields, etc. They do so with *low power* and *low mass* micro-instruments (on the order of milliwatts and grams) such as a Multispectral Microscopic Imager, a curved focal plane array camera for large field of view, a compact UV spectrometer, a visible spectrometer with no moving parts, a microweather station, etc. The moballs use wind for locomotion and generate *all* the electricity they need for their sensors and electronic parts from the natural resources on Mars (energy could be harvested from motion in the same way as automatic wristwatches, from sunlight using solar panels installed on the surface of the moball, and from diurnal temperature variations using thermoelectric generators). The moballs could be released using a *deployer* sent to Mars---think of pollination. This would be done at a correct altitude and with an appropriate velocity to ensure widespread and safe landing. Various strategies for deployment could be envisioned, such as sending multiple parachutes to several different geographic locations. The moballs communicate with each other and earth through a satellite system orbiting Mars. There could be also peer-to-peer communication between the moballs, creating a *network of shared data, computing, and tasks.*

GOWON is a 2012 Step B invitee for NASA Innovative Advanced Concept (NIAC). It addresses the challenge area of the *Mars Surface System Capabilities* area. We believe the challenge to be near-term, i.e., 2018-2024.

## Motivation and Rationale

Thanks to earlier exploration missions, we now have a much better understanding of the natural characteristics of Mars. We know that there is an abundance of wind at the surface (with average speeds of 10 m/s and much higher maximum speeds [1]), dust storms, high levels of saltation, diurnal (day/night) temperature swings of $100^{\circ}C$[2], etc. We have communication satellite systems in place that orbit Mars and that can monitor its surface. Future Mars missions must therefore attempt to *leverage* these characteristics and capabilities, and may do so by exploiting recent advances in low power micro-devices using MEMS (Micro-Electro-Mechanical System) technologies and others, miniature cameras, miniature wet chemistry labs, integrated circuits, low power wireless devices, etc. We believe the system proposed here addresses this opportunity head on.

## System Overview

Our proposed system is much more than a collection of sensors: the system is larger than the sum of its parts. In its entirety it is a *wireless mobile mesh network*.

While the moballs have peer-to-peer connections they could be also connected to the Mars orbiting satellite system, creating a harmonious network of shared data, computing, positioning, and tasks. The system constantly updates a global map of where the moballs are, where they are heading, and what each moball's energy and memory resources are. As a result, the moballs could cooperate in intelligent ways. For example, if several moballs sense that they are in close proximity to one another they need not all perform the same tasks; one moball could measure the local temperature, the other the local vapor content, etc., thereby saving the system power and reducing data traffic. On the other hand, if for some reason, highly reliable data is required, the moballs could fuse their measurements (by averaging, say) and send much more accurate values to the satellite and Earth. . The moballs should be aware of the topology of their neighbors and should also know how much battery power they and their neighbors have, how full their memory is, etc. This would allow them to intelligently prioritize and distribute tasks so as to minimize power consumption and data traffic.

### Risks and Challenges of the Approach

It is conceivable that a fraction of the moballs could get trapped in a closed area. We would perform thorough simulations, using Martian *circulation models and topology maps* to determine how to distribute the moballs to minimize the risk of this happening. Mars has very strong occasional wind gusts and we believe that these could easily free the light and bouncy moballs that could be stuck behind rocks or trapped in gullies. Even if a moball stays in a closed area for a long time, it could play the role of a fixed sensor. As long as it vibrates, uses its solar or thermoelectric mechanisms, it could generate the electricity it needs. We are always interested in having information from a fixed location on Mars over a long period of time.

Knowing where the moballs are is of great importance (one needs to know the location on Mars where a measurement was made). In the absence of GPS satellites on Mars we need to devise alternative solutions to be able to accurately locate the moballs. For example, a micro-gyroscope or the two-way communication links between various moballs could be used to measure the time of flight between different moballs. This information could be shared among various moballs to perform triangulation to create an internal and dynamically modifiable map of their locations.

### Additional Advantages

Current systems on Mars generate limited data from a single location (e.g. Phoenix), or from a very few locations (Opportunity, in its entire mission of 7.5 years traveled 33.5 km). Also they are not able to traverse and explore cluttered terrains, steep mountains or sand dunes as GOWON could. The current missions have limited life span since they need to rely on either RTG batteries (which would not last forever) or solar panels (which would not generate power when it is dark, winter, dusty, or in the higher latitudes). GOWON could generate much more *selective* and much more *robust* data from *many different types of terrains*. Since the moballs could be distributed across the surface of Mars and move around in a random fashion they have a much *higher chance of detecting rare and interesting phenomena, or areas with important samples for return to Earth*. Current Mars missions treat the Mars environment as hostile and try to overcome this hostility (such as dust-storms, steep slopes, cluttered surfaces, sand- dunes, winters, large temperature swings, etc.). GOWON, however, is envisioned to be Mars *compatible* and is envisioned to exploit the natural resources available on Mars. GOWON is envisioned as an *ongoing* and *scalable* mission. We could always upgrade the software in the moballs and satellites and, at the same time, add new moballs and satellites with new functionalities. Most current Mars missions are sophisticated devices with multiple functionalities. As a result, it is very difficult to recover from one of the functionalities going down. GOWON, however, is extremely *fault tolerant* and would work even if many sensors go down. Even compared to recent proposed mission concepts that share some similarities to what is proposed here, GOWON has many advantages. Unlike the single (20 kg, 6m) Tumbleweed [3] that would act as a single wind-driven rover, GOWON could perform intelligent distributed sensing across a whole network of lightweight sensors. Unlike a *fixed* sensor web such as Smart Dust [4], it could explore *different* areas. Being mobile, much smaller number of moballs could monitor a given area compared to what we would have needed with *fixed* sensors. Furthermore, one could expect that a fixed sensor network of small sensors may get covered and rendered inoperable following a Martian sandstorm; a mobile network of relatively large moballs would not be endangered by this. Finally, unlike the recently proposed Microbot[5] hoppers that are believed to have a 30-day life span, since they require large amounts of power for locomotion, GOWON could have a much longer lifespan.

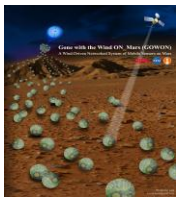

**References:** [1] M.I. Richardson, et al, "PlanetWRF: A General Purpose, Local to Global Numerical Model for Planetary Atmospheric and Climate Dynamics," J. Geophys., Res., 112, 2007. [2] H.H. Kieffer et al, "Thermal and albedo mapping of Mars during the primary Viking mission", J. Geophys. Res., 82; 4249-4291, 1977 [3] A. Behar et al, ↟"NASA/JPL Tumbleweed Polar Rover↡", Proceeding of the IEEE Aerospace Conference, March 2004 [4] B. Warnecke et al, ↟"Smart Dust: communicating with a cubic-millimeter compute↡", COMPUTER, vol 34, no 1, pages 44-51, 2001 [5] S. Dubowsky et al ↟"A Concept Mission: Microbots for Large-Scale Planetary Surface and Subsurface Exploration↡